\documentclass[AMA,STIX1COL]{WileyNJD-v2}
\usepackage{moreverb}
\usepackage{amsmath, stackrel, bm, bbm}
\usepackage{algorithm, algpseudocode, xcolor, subcaption, orcidlink} 

\usepackage{array}
\usepackage{colortbl} % For color control
\usepackage{arydshln} % For dashed lines

\newcommand\BibTeX{{\rmfamily B\kern-.05em \textsc{i\kern-.025em b}\kern-.08em
T\kern-.1667em\lower.7ex\hbox{E}\kern-.125emX}}

\articletype{RESEARCH ARTICLE}%

\usepackage{tikz,xcolor,hyperref}

\begin{document}

\title{Variable selection via fused sparse-group lasso penalized multi-state models incorporating molecular data}
% Short title: Variable selection for penalized multi-state models

\author[1,2]{Kaya Miah}

\author[3]{Jelle J. Goeman}

\author[3,4]{Hein Putter}

\author[1]{Annette Kopp-Schneider}

\author[1]{Axel Benner}

\authormark{Miah \textsc{et al}}

\address[1]{\orgdiv{Division of Biostatistics}, \orgname{German Cancer Research Center (DKFZ)}, \orgaddress{\state{Heidelberg}, \country{Germany}}}

\address[2]{\orgdiv{Medical Faculty}, \orgname{Heidelberg University}, \orgaddress{\state{Heidelberg}, \country{Germany}}}

\address[3]{\orgdiv{Department of Biomedical Data Sciences}, \orgname{Leiden University Medical Center (LUMC)}, \orgaddress{\state{Leiden}, \country{The Netherlands}}}

\address[4]{\orgdiv{Mathematical Institute}, \orgname{Leiden University}, \orgaddress{\state{Leiden}, \country{The Netherlands}}}

\corres{Kaya Miah, Division of Biostatistics, German Cancer Research Center (DKFZ), Im Neuenheimer Feld~280, D-69120 Heidelberg, Germany. \email{k.miah@dkfz.de}}

\presentaddress{Deutsche Forschungsgemeinschaft (DFG), Grant number: 514653984}

\abstract[Abstract]{In multi-state models based on high-dimensional data, effective modeling strategies are required to determine an optimal, ideally parsimonious model. 
In particular, linking covariate effects across transitions is needed to conduct joint variable selection. A useful technique to reduce model complexity is to address homogeneous covariate effects for distinct transitions. We integrate this approach to data-driven variable selection by extended regularization methods within multi-state model building. We propose the fused sparse-group lasso (FSGL) penalized Cox-type regression in the framework of multi-state models combining the penalization concepts of pairwise differences of covariate effects along with transition grouping. For optimization, we adapt the alternating direction method of multipliers (ADMM) algorithm to transition-specific hazards regression in the multi-state setting. In a simulation study and application to acute myeloid leukemia (AML) data, we evaluate the algorithm's ability to select a sparse model incorporating relevant transition-specific effects and similar cross-transition effects. We investigate settings in which the combined penalty is beneficial compared to global lasso regularization. 
}

%\keywords{Markov models; Regularization; Transition-specific hazards; Optimization algorithm; Survival analysis; Cox regression; High-dimensional data}

\maketitle

\section{Introduction}\label{sec:intro}
\begin{figure}[bt]
\centering
\includegraphics[width=17cm]{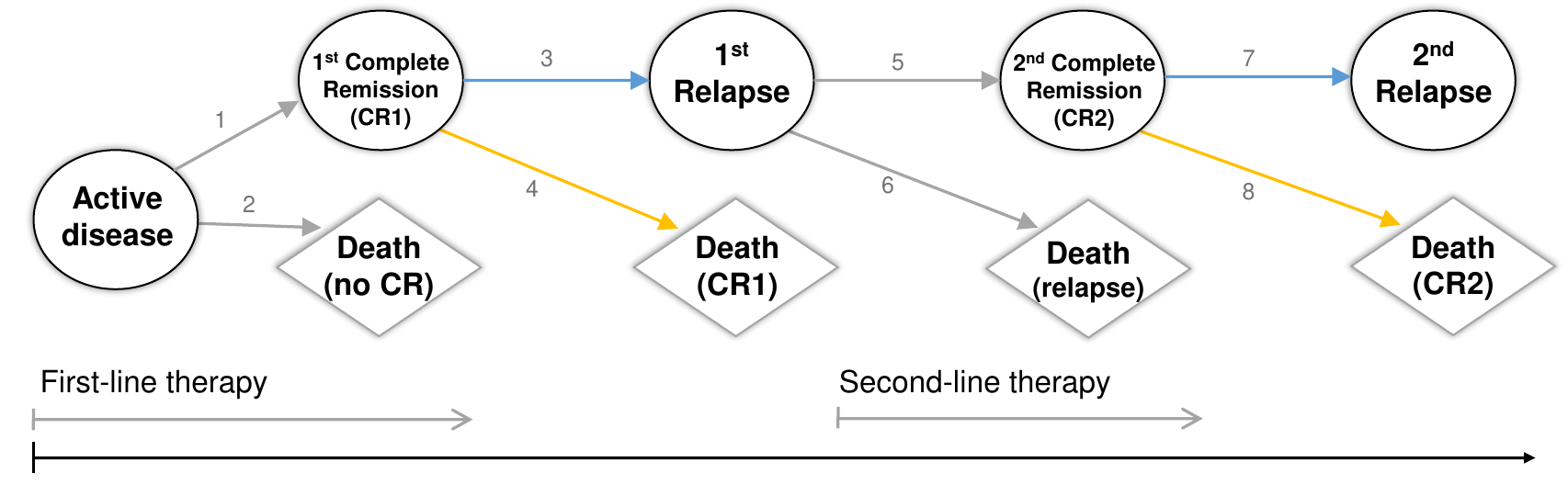}
\caption{State chart of the multi-state model for acute myeloid leukemia (AML) with nine states and eight possible transitions represented by arrows. Numbers denote the corresponding transition.}
\label{fig:diagram_AML}
\end{figure}

%Background
In medical research, common prediction models still predominantly make use of composite endpoints such as progression- or event-free survival. However, these time-to-first-event endpoints do not take into account important aspects of the individual disease and therapy course. Multi-state models are a natural framework to assess the effect of prognostic factors and treatment on the event history of a patient and to separate risks for the occurrence of distinct events. 
These extend competing risks analyses of event time endpoints such as time to progression, relapse, remission or death, by modeling the sequence of competing consecutive events on a macro level.
In survival analysis, the multi-state model class is used for event history data where individuals experience a sequence of events over time. Each event is defined by an entry and exit time along with transition types.
This paper is motivated by an application to the acute myeloid leukemia (AML) disease pathway. Figure~\ref{fig:diagram_AML} illustrates the event history for AML patients in the form of a state chart of a multi-state model with nine states and eight transitions. Distinct states are treated as nodes and possible transitions are represented by directed arrows.
To assess how probabilities of going from state to state depend on covariates, multi-state proportional hazards regression models can be used. In the era of precision medicine with increasingly high-dimensional information on molecular biomarkers, such a holistic analysis of a multi-state model is of essential interest. For our motivating AML application, we will investigate the effect of biomarkers along with established clinical covariates on the different transitions of the multi-state model depicted in Figure~\ref{fig:diagram_AML}.
Thus, effective variable selection strategies for multi-state models incorporating high-dimensional molecular data are required to obtain a sparse model and mitigate overfitting. Such data-driven model building strategies will contribute to a deeper understanding of the individual disease progression and its therapeutic concepts as well as improved personalized prognoses. \\

%Research question
This paper focuses on data-driven variable selection via penalized multi-state models to capture pathogenic processes and underlying etiologies more accurately. The goal is to select a sparse model based on high-dimensional molecular data by extended regularization methods. 
We want to use a-priori knowledge about the multi-state model structure to help simplify it. First, we assume that most biomarkers might have no effect on specific disease transitions, and if they have an effect on one transition, they might have an effect on many. Second, we presume that parameters of similar transitions are often of a similar direction and magnitude. Third, biomarker effects might only be relevant for specific transitions. Thus, the parameter space dimensionality should be decreased by setting non-relevant biomarker effects to zero (i.e. sparsity), identifying similar biomarker effects across distinct transitions (i.e. similarity) and detecting only relevant biomarker effects for specific transitions of interest (i.e. group structuring). Further, we want to combine molecular and clinical data by incorporating established clinical predictors into the model that remain unpenalized. In our motivating AML application illustrated in Figure~\ref{fig:diagram_AML}, we assume that biomarker effects of transitions 3 and 7, i.e. from first complete remission (CR1) to first relapse and from second complete remission (CR2) to second relapse, might be similar. Further, we would not expect any biomarker effect on e.g. transition 1, i.e. from active disease to early death, since this might rather be related to the initiation of intensive chemotherapy. \\

%%%%%%%%%%%%%%%%%%%%%%%%%%%%%%%%%%%%%%%%%%%%%%%%%%%%%%%%%%%%%%%%%%%%%%%%%%%%%%%%%%%%%%%%%%%%%%%%%%%%%%%%%%%%%%%%%%%%%%
% SCOPING REVIEW OVERVIEW
A scoping literature review on statistical methods for model selection in the framework of multi-state models was conducted based on the PubMed database (\texttt{\url{http://www.ncbi.nlm.nih.gov/pubmed/advanced}}).
In the following, we give a brief overview on existing model selection strategies for multi-state model building in survival analysis.

Common methods for variable selection comprise regularization in the fitting process in order to avoid the inclusion of covariates with non-relevant effects. 
%We focus on the Cox-type proportional hazards model adapted on transition-specific hazards for multi-state outcomes.
\citet{saadati_prediction_2018} proposed a lasso penalized cause-specific hazards approach for competing risks data in higher dimensions, where the independently penalized cause-specific hazards models are linked by choosing the combination of tuning parameters that yields the best prediction.
In the multi-state setting, \citet{sennhenn-reulen_structured_2016} developed a data-driven regularization method for sparse modeling by combining so-called cross-transition effects of the same baseline covariate. The structured fusion lasso penalization regularizes the $L_1$-norm of the regression coefficients as well as all pairwise differences between distinct transitions.
\citet{dang2021risk} suggested $L_1$-penalization by a one-step coordinate descent algorithm. Further, \citet{huang_elasticnet_2018} proposed a regularized continuous-time Markov model with the elastic net penalty. 
%
%Especially in higher dimensions, statistical boosting algorithms are powerful techniques with respect to model selection. 
%A scoping review on statistical boosting algorithms in biomedical research is given in \citet{mayr_update_2017}. 
Beyond, \citet{reulen_boosting_2016} introduced the component-wise functional gradient descent boosting approach to perform unsupervised variable selection and multi-state model choice simultaneously. In particular, they focused on non-linearity of single transition-specific or cross-transition effects. 
Further, \citet{edelmann_global_2020} extended the global test to competing risks and multi-state models to test if regression coefficients for a certain subset of transitions are equal under the Markov assumption.
\citet{fiocco_reduced_2005} introduced reduced rank proportional hazards regression to competing risks and multi-state models\citep{fiocco_reduced-rank_2008} for limiting the number of regression parameters. 
A model class to directly estimate the effect of a covariate on survival times are accelerated failure time (AFT) models. \citet{huang_two-sample_2000} established the multi-state accelerated sojourn times model. \citet{ramchandani_estimation_2020} yield insights into the estimation of an AFT model with intermediate states as auxiliary information.  
With respect to model selection approaches, \citet{huang2006regularized} consider regularized AFT models with high-dimensional covariates.
Beyond, pseudo-observations in event history analysis introduced by \citet{andersen_generalised_2003} provide another direct modeling technique. The state occupation probabilities are modeled directly instead of considering each transition intensity separately. These pseudo-values are then used in a generalized estimating equation (GEE) to deduce estimates of the model parameters. In terms of variable selection procedures, \citet{wang2012penalized} proposed penalized GEE based on high-dimensional longitudinal data. Further, \citet{niu_variable_2020} utilize penalized GEE for a marginal survival model. Based on pseudo-observations, \citet{su_analysis_2022} make use of penalized GEE for proportional hazards mixture cure models. 
We focus on the well established hazard-based framework of Cox-type multi-state models, so that direct modeling approaches are not further pursued in this work. 
Consequently, there is already some valuable work for multi-state model selection. But to the best of our knowledge, a-priori information on the multi-state model structure is not yet well taken into account. \\

%%%%%%%%%%%%%%%%%%%%%%%%%%%%%%%%%%%%%%%%%%%%%%%%%%%%%%%%%%%%%%%%%%%%%%%%%%%%%%%%%%%%%%%%%%%%%%%%%%%%%%%%%%%%%%%%%%%%%%

%Solution
In this paper, we propose the fused sparse-group lasso (FSGL) penalty for multi-state models combining the concepts of general sparsity, pairwise differences of covariate effects and transition grouping. For fitting such a penalized multi-state model, we adapt the alternating direction method of multipliers (ADMM) optimization algorithm to Cox-type hazards regression in the multi-state setting due to its beneficial feature of decomposing the objective function. \\

%Outline
The remainder of the paper is structured as follows: The methodological background of multi-state models needed for the proposed adaptation is given in Section~\ref{sec:multistate_models}. Section~\ref{sec:FSGL} introduces the FSGL penalty extended to the multi-state setting. Section~\ref{sec:Optimization} describes the general ADMM optimization algorithm for parameter estimation in Subsection~\ref{sec:ADMM} along with the derived ADMM update steps to fit FSGL penalized multi-state models in Subsection~\ref{sec:ADMM-mstate}.  Section~\ref{sec:simulations} shows the results of a proof-of-concept simulation study to investigate the regularization performance of the derived algorithm and Section~\ref{sec:application} illustrates a real data application to AML patients. 
%Section~\ref{sec:discussion} summarizes with concluding remarks.

%%%%%%%%%%%%%%%%%%%%%%%%%%%%%%%%%%%%%%%%%%

\section{Methods for multi-state modeling}\label{sec:multistate_models}

The following Section provides a brief introduction to the multi-state model class in survival analysis needed for our adapted fused sparse-group lasso penalty to the multi-state setting. A holistic framework to multi-state modeling theory can be found in \citet{andersen_1993}. Subsection~\ref{sec:cs_Cox} introduces the general multi-state process and defines the concept of transition-specific Cox proportional hazards regression for multi-state models. Further, Subsection~\ref{sec:Likelihood} denotes the explicit likelihood formulation in the multi-state setting along with its derivatives needed for model fitting.

\subsection{Multi-state proportional hazards regression model}\label{sec:cs_Cox}

Following \citet{andersen_multi-state_2002} and \citet{Putter_mstate_2007}, a multi-state process is a stochastic process $\lbrace Z(t), t \in \mathcal{T}\rbrace$ with times in $\mathcal{T} = [0,t_{\text{max}}], 0 < t_{\text{max}} < \infty$, and a finite state space $\mathcal{K} = \lbrace 1,\dots,K \rbrace$. The transition probabilities are given as
\begin{align*}
    \text{P}_q(s,t) = \text{P}_{[k.k']}(s,t) = \text{P}(Z(t) = k' ~|~ Z(s) = k)
\end{align*}
for transition $q = [k.k']$ from state $k$ to $k', k, k' \in \mathcal{K}, s,t \in \mathcal{T}, s \leq t$ and $q \in \mathcal{Q} = \lbrace 1,\dots,Q\rbrace$ the set of observable transitions. We assume a Markovian model, i.e. the probability for a transition only depends on the current state of the multi-state process at the current time.
The transition intensities are defined as the corresponding derivatives
\begin{align*}
    h_q(t) = \lim_{\Delta t \searrow 0} \frac{\text{P}_q(t,t+\Delta t)}{\Delta t} .
\end{align*}

To assess the dependence on covariates, these transition-specific hazard rates can be modeled by separate Cox proportional hazards models for each transition as
\begin{align*}
    h_q (t|\bm{x}) = h_{0,q}(t) \exp \lbrace  \bm{\beta}_q^T \bm{x} \rbrace , ~ q=1,\dots,Q,  %\label{eq:hazard-model}
\end{align*}
for an individual with covariate vector $\bm{x}=(x_{1},\dots,x_{P})^T \in \mathbb{R}^P$, where $h_{0,q}(t)$ denotes the baseline hazard rate of transition $q$ at time $t$ and $\bm{\beta}_q = (\beta_{1,q},\dots, \beta_{P,q})^T \in \mathbb{R}^P$ the vector of transition-specific regression coefficients. Thus, Cox-type regression analysis for multi-state data enables simultaneous modeling of the relationship between covariates and all relevant transitions.\cite{rademacher2022}

\subsection{Multi-state likelihood formulation}\label{sec:Likelihood}

In the multi-state framework, the generalized partial likelihood can be written in terms of a stratified formulation as a product of Cox partial likelihoods for each transition, i.e.
\begin{align*}
    l(\bm{\beta}) = \prod_{q=1}^Q l_q(\bm{\beta}_q) 
    = \prod_{q=1}^Q ~\prod_{i=1}^N \left( \dfrac{\exp \lbrace \bm{x}_i^T \bm{\beta}_{q} \rbrace}{\sum_{l \in R_{i,q}} \exp \lbrace \bm{x}_l^T \bm{\beta}_{q} \rbrace} \right) ^{\delta_{i,q}}, 
    %= \prod_{i=1}^N \left[ \prod_{q=1}^Q \left( \exp \left\lbrace - \int_0^{t_{max,i}} h_{q,i}(t) R_{q,i}(t) dt \right\rbrace \prod_{c=1}^{C_{q,i}(t_{max,i})} h_{q,i}(t_{q,i,c}) \right) \right]
\end{align*}
where $\bm{x}_i = (x_{1;i},\dots,x_{P;i})^T  \in \mathbb{R}^{P}$ denotes the covariate vector of individual $i, i=1,\dots,N$, $\beta_{q} \in \mathbb{R}^{P}$ the transition-specific regression vector, and $\delta_{i,q}$ the event indicator for transition $q$.\citep{Putter_mstate_2007, putter2006estimation}
$R_{i,q}$ denotes the risk set for individual $i$ with transition $q$ at time $t_i$. This set includes all individuals who are at risk of experiencing a transition of type $q$ transition at time $t_i$.
The transition-specific Cox partial likelihood $l_q(\beta_{q})$ compares the hazard of the individual with an event at time $t_i$ to the hazard of all individuals under risk at $t_i$.

The multi-state partial likelihood formulation for the stacked regression vector $\bm{\beta} = (\beta_{1,1},\dots,\beta_{1,Q}, \beta_{2,1},\dots,\beta_{P,Q})^T \in \mathbb{R}^{PQ}$ and corresponding extended covariate vector $\Tilde{\bm{x}}_i = (x_{1.1;i},\dots,x_{1.Q;i},x_{2.1;i},\dots,x_{P.Q;i})^T \in \mathbb{R}^{PQ}$ is then derived as
\begin{align*}
l(\bm{\beta}) = \prod_{i=1}^n \left( \dfrac{\exp \lbrace \Tilde{\bm{x}}_i^T \bm{\beta} \rbrace}{\sum_{l \in \Tilde{R}_i} \exp \lbrace \Tilde{\bm{x}}_l^T \bm{\beta} \rbrace} \right) ^{\delta_i}, 
\end{align*}
where $\Tilde{R}_i$ denotes the corresponding risk set formulation based on long format data according to \citet{wreede2010mstate} with single lines $i,j$ from a total of $n$ rows. In this format, each individual has a row for each transition for which they are at risk.
The negative logarithm of the multi-state partial likelihood is
\begin{align}
L(\bm{\beta}) = -\log[l(\bm{\beta})] = \sum_{i=1}^n \delta_i \left[ -\Tilde{\bm{x}}_i^T \bm{\beta} + \log \left(\sum_{l \in \Tilde{R}_i} \exp \lbrace \Tilde{\bm{x}}_l^T \bm{\beta} \rbrace \right) \right]. \label{eq:LL_mstate_long}
\end{align}

The regression parameters are then estimated by minimizing this negative partial log-likelihood. The estimate $\hat{\bm{\beta}}$ is plugged-in in Breslow's estimate of the cumulative baseline hazard\citep{Putter_mstate_2007}
\begin{align*}
    \widehat{\Lambda}_{0,q}(t) = \sum_{j: t_j \leq t} \frac{1}{\sum_{l \in R_{j,q}} \exp \lbrace \bm{x}_l^T \hat{\bm{\beta}}_q \rbrace}.
\end{align*}
For estimation, we need the first and second derivative of the Cox partial log-likelihood function. The score vector is given as 
\begin{align}
    U(\bm{\beta}) = \frac{\partial}{\partial \beta} \log[l(\bm{\beta})] = \bm{X}^T (\bm{\delta}-\hat{\bm{\mu}}),
    \label{eq:LL-score}
\end{align}
where $\bm{X} \in \mathbb{R}^{n \times PQ}$ denotes the design matrix, $\bm{\delta} = (\delta_1,\dots,\delta_n)^T$ the vector of event indicators and $\hat{\bm{\mu}} = (\hat{\mu}_1, \dots,\hat{\mu}_n)^T$ the estimated cumulative hazards with elements $\hat{\mu}_i~=~\widehat{\Lambda}_0(t_i) \exp \lbrace \bm{x}_i^T \bm{\beta} \rbrace$. The Hessian matrix is
\begin{align}
    J(\bm{\beta}) = \frac{\partial^2}{\partial \beta \partial \beta^T} \log[l(\bm{\beta})] = -\bm{X}^T \bm{W} \bm{X} 
    \label{eq:LL-hessian}
\end{align}
with $\bm{W} \in \mathbb{R}^{n \times n}$ the weight matrix of the estimated cumulative hazards $\hat{\bm{\mu}}$.\citep{goeman2010, van_houwelingen_cross-validated_2006}

%%%%%%%%%%%%%%%%%%%%%%%%%%%%%%%%%%%%%%%%%%

\section{Fused sparse-group lasso penalty}\label{sec:FSGL}

This Section describes our adapted fused sparse-group lasso (FSGL) penalty to multi-state models as key variable selection strategy for high-dimensional multi-state modeling. \\

For data-driven model selection, established methods incorporate regularization in the fitting process in order to conduct variable selection.\citep{benner_high-dimensional_2010, heinze_variable_2018} Especially in applications with few events per variable, regularization is needed in order to obtain a unique and more stable solution of the regression parameters.\citep{salerno2023high_review}
Several regularization methods that perform covariate selection beyond the least absolute shrinkage and selection operator (lasso) \citep{tibshirani1996} have been developed. 
These include elastic net \citep{zou_regularization_2005}, fused lasso \citep{tibshirani2005fused}, sparse-group lasso \citep{simon_sparse-group_2013} and fused sparse-group lasso \citep{zhou2012fsgl} penalization.
In the multi-state framework, adapted regularization approaches incorporate the lasso \citep{saadati_prediction_2018, dang2021risk}, elastic net \citep{huang_elasticnet_2018} and structured fusion lasso \citep{sennhenn-reulen_structured_2016} for penalized multi-state modeling. Table~\ref{tab:penalties} gives an overview of existing penalization methods along with their penalty functions as well as their original publications for linear regression models and adaptations to Cox models for survival outcomes. \\

%All described selection methods are based on the Cox proportional hazards model adapted on transition-specific hazards for time-to-event outcomes. 

 \begin{table}[b]
\centering
\caption{Examples of penalization methods.}
\label{tab:penalties}
\begin{tabularx}{\textwidth}{llll}\hline
\toprule
\textbf{Penalization method} & \textbf{Penalty function} & \textbf{Parameters} & \textbf{Model type} \\ \hline\hline
Ridge & $\lambda \| \bm{\beta} \|_2^2$ & $\lambda > 0$ & Linear (\citet{hoerl1970ridge}), \\
 & & & Cox (\citet{gray1992ridgecox},  \\
 &&& \citet{verweij1994penalized}) \\
Lasso  &  $\lambda \| \bm{\beta} \|_1$ & $\lambda > 0$ & Linear (\citet{tibshirani1996}),\\
 & & & Cox (\citet{tibshirani1997lassocox}) \\
Elastic net  & $\alpha \| \bm{\beta} \|_1 + (1-\alpha) \| \bm{\beta} \|_2^2$ & $ \alpha \in [0,1]$ & Linear (\citet{zou_regularization_2005}), \\
 & & & Cox (\citet{simon2011regularization}) \\
Fused lasso  & $\lambda_1 \sum_{p=1}^P |\beta_p| + \lambda_2 \sum_{p=2}^P |\beta_p - \beta_{p-1}|$ & $\lambda_1, \lambda_2 >0$ & Linear (\citet{tibshirani2005fused}),\\
 & & & Cox (\citet{chaturvedi2014fused}) \\
Group lasso & $\lambda \sum_{g\in \mathcal{G}} \sqrt{p_g} \| \bm{\beta}_g \|_2$ & $\lambda>0$, groups $\mathcal{G}$,  & Linear (\citet{yuan2006}), \\
 & & group size $p_g$ & Cox (\citet{kim2012group})\\
Sparse-group lasso  & $\alpha \| \bm{\beta} \|_1 + (1-\alpha) \sum_{g\in \mathcal{G}} \sqrt{p_g} \| \bm{\beta}_g \|_2$ &  $\alpha \in [0,1]$ & Linear \& Cox (\citet{simon_sparse-group_2013})\\
Fused sparse-group lasso  & $\lambda ~[ \alpha \gamma \|\bm{\beta}\|_1 + (1-\gamma) \|\bm{D} \bm{\beta}\|_1$ & $\lambda >0, \alpha, \gamma \in [0,1]$, & Linear (\citet{beer_incorporating_2019})\\
 & $+ (1-\alpha)\gamma \sum_{g\in \mathcal{G}} \sqrt{p_g} \| \bm{\beta}_g \|_2 ]$ & fusion matrix $\bm{D}$ \\
 
\hline
Lasso mstate  & $\lambda \sum_q \sum_p |\beta_{p,q}|$ & $\lambda > 0$ & Competing risks (\citet{saadati_prediction_2018}), \\
&&& Multi-state (\citet{dang2021risk})\\
Elastic net mstate & $(1-\alpha)  \sum_{p,q} \beta_{p,q}^2 + \alpha \sum_{p,q} |\beta_{p,q}|$ &  $\alpha \in [0,1]$ & Multi-state (\citet{huang_elasticnet_2018})\\
Fusion lasso mstate  & $\lambda_1 \sum_q \sum_p |\beta_{p,q}| $ & $\lambda_1, \lambda_2 >0$ & Multi-state \\
 & $+ \lambda_2 \sum_{q,q'} \sum_{p=1}^P |\beta_{p,q} - \beta_{p,q'}|$ & & (\citet{sennhenn-reulen_structured_2016}) \\
\bottomrule
\hline
\end{tabularx}
\end{table}

The \textit{fused sparse-group lasso} (FSGL) penalty, introduced by \citet{zhou2012fsgl} and adapted by \citet{beer_incorporating_2019} for linear models, provides a combination of lasso, fused and grouped regularization. Thus, prior information of spatial and group structure can be incorporated into the prediction model. The global lasso penalty fosters overall sparsity. The fusion penalty regularizes absolute pairwise differences of regression coefficients. The group penalty allows variables within the same group to be jointly selected or shrunk to zero.

We propose to transfer this combined penalty to the multi-state framework based on transition-specific hazards regression models in order to obtain overall sparsity, link covariate effects across transitions and incorporate transition grouping. Thus, we advocate the FSGL penalty that provide estimates with three properties:
\begin{enumerate}
\itemsep-0.3em 
    \item \textit{Sparsity}: The resulting estimator automatically zeros out small estimated coefficients to achieve variable selection and simplify the model.\citep{fanli2002}
    \item \textit{Similarity}: The resulting estimator penalizes absolute differences of covariate effects across similar transitions, thus addressing homogeneous cross-transition effects.
    \item \textit{Group structuring}: The resulting estimator allows variables within the same transition to be jointly selected or shrunk to zero, thus incorporating transition grouping. 
\end{enumerate}

We consider the same set of $P$ (time-fixed) covariates, e.g. biomarkers, for each transition $q \in \lbrace1,\dots,Q \rbrace = \mathcal{Q}$. Further, we presume a subset of pairs of similar transitions $\mathcal{S} = \lbrace (q,q'): q \neq q', q,q' \in \mathcal{Q} \rbrace$, of which we assume that covariate effects across these transitions are of a similar magnitude, i.e. we consider potential cross-transition effects.\citep{sennhenn-reulen_structured_2016} The FSGL penalty function is then defined as
\begin{align}
p_{\lambda, \text{FSGL}}(\bm{\beta}) &= \lambda \left[ \alpha \gamma \sum_{q=1}^Q \sum_{p=1}^P |\beta_{p,q}| 
+ (1-\gamma)\sum_{(q,q') \in \mathcal{S}} \sum_{p=1}^P |\beta_{p,q} - \beta_{p,q'}| 
+ (1-\alpha)\gamma \sum_{q=1}^Q \|\bm{\beta}_{q}\|_2 \right] , \label{eq:FSGL-penalty}
\end{align}
with transition-specific regression coefficients $\beta_{p,q}$ of covariate $x_p, p=1,\dots,P$ for transition $q$, transition-specific regression vector $\bm{\beta}_{q}~\in~\mathbb{R}^{P}$ and tuning parameters $\lambda, \alpha, \gamma$. 
The tuning parameter $\lambda >0$ controls the overall level of regularization, $\alpha \in [0,1]$ balances between global lasso and group lasso and $\gamma \in [0,1]$ balances between sparse penalties and the fusion penalty.\citep{beer_incorporating_2019}
Thus, the optimal tuning parameter $\lambda_{\text{opt}}$ is chosen at pre-selected values of $\alpha$ and $\gamma$. For $(\alpha, \gamma) = (1,1)$, the estimator reduces to the global lasso, for $(\alpha, \gamma) = (0,1)$ to the group penalty and for $(\alpha, \gamma) = (1,0)$ or $(\alpha, \gamma) = (0,0)$ to the fusion penalty.
The regression parameter $\beta$ is estimated by minimizing the penalized negative partial log-likelihood function, i.e.
\begin{align*}
    \hat{\bm{\beta}} = \text{arg min}_\beta \left[ L(\bm{\beta}) + p_{\lambda, \text{FSGL}} (\bm{\beta}) \right].
\end{align*}

%%%%%%%%%%%%%%%%%%%%%%%%%%%%%%%%%%%%%%%%%%
\section{Optimization algorithm}\label{sec:Optimization}

This Section introduces the general concept of the alternating direction method of multipliers (ADMM) optimization algorithm in Subsection~\ref{sec:ADMM} and provides the explicitly derived ADMM updating steps to fit FSGL penalized multi-state models in Subsection~\ref{sec:ADMM-mstate}. The criterion of selecting optimal penalty parameters is described in Subsection~\ref{sec:Tuning}.\\

For penalized Cox-type regression, several numerical optimization algorithms exist for parameter estimation by minimizing the penalized negative likelihood function. 
\citet{simon_sparse-group_2013} utilize an accelerated generalized gradient algorithm for the sparse-group lasso penalty. However, the accelerated gradient method depends on the separability of the penalty term across groups of $\bm{\beta}$, so that the fusion penalty can only be applied within groups.
For the structured fusion lasso penalty, \citet{sennhenn-reulen_structured_2016} make use of a penalized iteratively re-weighted least squares algorithm. This second-order optimization has high computation cost and potential convergence problems.\citep{dang2021risk}
Further, coordinate descent algorithms do not work for the fused lasso penalty due to its non-separability into a sum of functions of the elements of $\bm{\beta}$ that is also not continuously differentiable.
Thus, we chose the ADMM optimization algorithm to fit FSGL penalized multi-state models due to the decomposability of the objective function as well as superior convergence properties.

%%%%%%%%%%%%%%%%%%%%%%%%%%%%%%%%%%%%%%%%%%
\subsection{Alternating direction method of multipliers algorithm}\label{sec:ADMM}

The \textit{alternating direction method of multipliers} (ADMM) algorithm provides a very general framework for numerical optimization of convex functions. It originates from the 1950s and was developed in the 1970s\citep{gabay1976dual, glowinski1975}, but was holistically examined later by \citet{boyd2010} for a broader conceptuality. 
The algorithm combines the decomposability of the objective function with superior convergence properties of the method of multipliers.\citep{boyd2010}
Consider the following general optimization problem w.r.t. a variable $\bm{\beta} \in \mathbb{R}^P$ 
\begin{align*}
    \text{min}_\beta ~ f(\bm{\beta}) + g(\bm{\beta}),
\end{align*}
where $f,g$ denote convex functions.
In the ADMM framework, the generic constrained optimization problem introducing an auxiliary variable $\bm{\theta} \in \mathbb{R}^P$ is given as
\begin{align*}
    \text{min}_{\beta,\theta} ~ f(\bm{\beta}) + g(\bm{\theta}) ~~~~\text{subject to}~~ \bm{\theta} - \bm{\beta} = \bm{0} .
\end{align*}
Thus, the objective function becomes additively separable, which simplifies the subsequent optimization steps.
As in the method of multipliers, the augmented Lagrangian function adding an $L_2$-term to enhance optimization stability\cite{parka2022admm} is given as
    \begin{align*}
        &\mathcal{L}(\bm{\beta}, \bm{\theta}, \bm{\phi}) = f(\bm{\beta}) + g(\bm{\theta}) + \bm{\phi}^T (\bm{\theta} - \bm{\beta}) + \frac{\rho}{2} \|\bm{\theta} - \bm{\beta}\|_2^2 \\ 
        =~&\mathcal{L}(\bm{\beta}, \bm{\theta}, \bm{\nu}) = f(\bm{\beta}) + g(\bm{\theta}) + \frac{\rho}{2} \|\bm{\theta} - \bm{\beta} + \bm{\nu} \|_2^2 - \frac{\rho}{2} \| \bm{\nu} \|^2 ,
    \end{align*}
with Lagrangian multiplier $\bm{\phi} \in \mathbb{R}^P$, augmented Lagrangian parameter $\rho >0$ (i.e. the ADMM step size) and scaled dual variable $\bm{\nu} = \frac{\bm{\phi}}{\rho} \in \mathbb{R}^P$.
The general ADMM iterations consist of the following alternating update steps at iteration $r+1$: 
    \begin{align*}
    \bm{\beta}^{r+1} &= \text{arg min}_{\beta} ~ \mathcal{L}(\bm{\beta}, \bm{\theta}^r, \bm{\nu}^r), \\
    \bm{\theta}^{r+1} &= \text{arg min}_{\theta} ~ \mathcal{L}(\bm{\beta}^{r+1}, \bm{\theta}, \bm{\nu}^r) ,\\
    \bm{\nu}^{r+1} &= \bm{\nu}^r + \bm{\beta}^{r+1} - \bm{\theta}^{r+1} .
    \end{align*}

The algorithm comprises a $\bm{\beta}$-minimization step, a $\bm{\theta}$-minimization step and a dual variable $\bm{\nu}$-update. Thus, the usual joint minimization is separated across the decomposition of the objective function over parameters $\bm{\beta}$ (e.g. likelihood) and $\bm{\theta}$ (e.g. penalty) into two steps. \\

As a stopping criterion, \citet{boyd2010} propose sufficiently small primal and dual residuals, i.e.
\begin{align*}
& \|\bm{\beta}^{r+1}-\bm{\theta}^{r+1}\|_2 < \epsilon_1 = \sqrt{P} \epsilon_{\text{abs}} + \epsilon_{\text{rel}} \max \lbrace \|\bm{\beta}^r\|_2,\|\bm{\theta}^r\|_2 \rbrace ~\text{and} \\
& \|\rho (\bm{\theta}^{r+1}-\bm{\theta}^{r})\|_2 < \epsilon_2 = \sqrt{P} \epsilon_{\text{abs}} + \epsilon_{\text{rel}} \|\nu^r\|_2 ,
\end{align*}
with absolute and relative tolerances $\epsilon_{\text{abs}} = 10^{-4}$ and $\epsilon_{\text{rel}} = 10^{-2}$.

%%%%%%%%%%%%%%%%%%%%%%%%%%%%%%%%%%%%%%%%%%%%
\subsection{ADMM for penalized multi-state models}\label{sec:ADMM-mstate}

In the FSGL penalized multi-state framework, the constrained optimization problem for the stacked regression parameter $\bm{\beta}~\in~\mathbb{R}^{PQ}$ is given as
\begin{align*}
    \text{min}_{\bm{\beta},\bm{\theta}}~ f(\bm{\beta}) + g(\bm{\theta}) ~~\text{subject to}~~ \theta_m - \bm{K}_m \bm{\beta} = 0, ~m \in \{1,\dots,M\},
\end{align*}
where $f(\bm{\beta}) = L(\bm{\beta})$ is the negative multi-state partial log-likelihood function as in (\ref{eq:LL_mstate_long}) and $g(\bm{\theta}) = p_{\lambda, \text{FSGL}}(\bm{\theta})$ is the FSGL penalty function (\ref{eq:FSGL-penalty}) with auxiliary variable $\bm{\theta} = (\theta_1,\dots,\theta_M)^T~\in~\mathbb{R}^M$, $M = PQ + s + PQ$, such that $\theta_m = \bm{K}_m \bm{\beta}$. The penalty structure matrix is defined as $\bm{K}~=~(\bm{K}_1|\dots|\bm{K}_M)^T~\in~\mathbb{R}^{M \times PQ}$, with elements $k_{ij} \in \lbrace -1,0,1 \rbrace$, such that
\[
\bm{K}_m = \left\{
   \begin{array}{ll}
      \bm{u}_m, & \text{if}~ m \in \lbrace 1,\dots,PQ \rbrace, \\
      \bm{d}_{m-PQ}, & \text{if}~ m \in \lbrace PQ+1,\dots,PQ+s \rbrace, \\
      %\bm{G}_1, & \text{if}~ m \in \lbrace PQ+s+1,\dots,PQ+s+P \rbrace, \\
      \bm{G}_{m-PQ-s}, & \text{if}~ m \in \lbrace PQ+s+1,\dots,PQ+s+Q \rbrace, \\
   \end{array}
\right.
\]
where $\bm{u}_m$ denotes the unit vector of the identity matrix $\bm{I}_{PQ} \in \mathbb{R}^{PQ \times PQ}$ corresponding to the global lasso penalty. The contrast vector of the $(m - PQ)$-th row of the fusion matrix $\bm{D} \in \mathbb{R}^{s \times PQ}$ for $s$ pairs of similar transitions with elements $d_{ij}~\in~\lbrace-1,1\rbrace$ at the corresponding positions of covariates of such similar transitions corresponding to the fusion penalty is denoted as $\bm{d}_m$, e.g. $\bm{d}_1 = (1, -1, 0, \dots, 0)^T$ for covariate $X1.1$ and $X1.2$ of transitions 1 and 2. $\bm{G}_{m-PQ-s} \in \mathbb{R}^{P \times PQ}$ are the group matrices of the $Q$ transitions consisting of unit vectors that indicate the group allocation of a variable to a corresponding transition for the group penalty.
The penalty structure matrix $\bm{K}$ is then given as
\begin{align*}
\bm{K} = 
\left[
    \begin{array}{ccc}
\bm{I}_{PQ} \\
\hdashline[4pt/2pt] \\[-1em]
\bm{D} \\
\hdashline[4pt/2pt] \\[-1em]
\bm{G}_1 \\
\vdots \\
\bm{G}_Q
    \end{array}
\right]
= 
\left[
    \begin{array}{cccccccc}
1 & 0 & 0 & \cdots & \cdots & 0 & 0 & 0 \\
0 & 1 & 0 & \cdots & \cdots & 0 & 0 & 0 \\
\vdots & \vdots & \vdots & \vdots & \vdots & \vdots & \vdots & \vdots \\
0 & 0 & 0 & \cdots & \cdots & 0 & 0 & 1 \\ 
\hdashline[4pt/2pt] \\[-1em]
1 & -1 & 0 & \cdots & \cdots & 0 & 0 & 0 \\
1 & 0 & -1 & \cdots & \cdots & 0 & 0 & 0 \\
\vdots & \vdots & \vdots & \vdots & \vdots & \vdots & \vdots & \vdots \\
0 & 0 & 0 & \cdots & \cdots & 0 & 1 & -1 \\ 
\hdashline[4pt/2pt] \\[-1em]
1 & 0 & 0 & \cdots & \cdots & 0 & 0 & 0 \\
0 & 0 & 0 & \cdots & \cdots & 1 & 0 & 0 \\
\vdots & \vdots & \vdots & \vdots & \vdots & \vdots & \vdots & \vdots \\
0 & 0 & 0 & \cdots & \cdots & 0 & 0  & 1
    \end{array}
\right]. 
\end{align*}
Thus, the total number of rows of the penalty structure matix $\bm{K}~\in~\mathbb{R}^{M \times PQ}$ is $M = PQ + s + PQ$. Optimization of the likelihood and penalty terms are separated and therefore simplified. \\

For the $\bm{\beta}$-updating step, Cox estimation of the regression parameter $\bm{\beta}$ is performed by numerical algorithms. The gradient descent update is given as $\bm{\beta}_{\text{GD}}^{r+1}~=~\bm{\beta}^r + \epsilon_{\text{GD}} U(\bm{\beta}^r)$ using the score vector $U(\bm{\beta}^r)$ at iteration $r$ as in (\ref{eq:LL-score}) and step size $\epsilon_{\text{GD}}$. The Newton-Raphson update is
\[
\bm{\beta}_{\text{NR}}^{r+1}~=~\bm{\beta}^r~-~J(\bm{\beta}^r)^{-1} U(\bm{\beta}^r),
\]
using both the gradient $U(\bm{\beta}^r)$ and Hessian matrix $J(\bm{\beta}^r)$ at iteration $r$ as in (\ref{eq:LL-hessian}). The estimation tolerance for the convergence criterion based on the partial log-likelihood is denoted as $v_{\text{NR}}$.
A hybrid algorithm as proposed by \citet{goeman2010} combines adaptive gradient descent and Newton-Raphson to derive $\bm{\beta}$-estimates in a Cox model. It starts with a single gradient descent step and then switches to Newton-Raphson updating steps. 
For an efficient $\bm{\theta}$-updating step, the proximity operator is utilized, i.e. the vector soft-thresholding operator for $\bm{a} \in \mathbb{R}^m$
\[
S_\kappa(\bm{a}) = (1 - \kappa/\|\bm{a}\|_2)_+ \bm{a} ,
\]
with $S_\kappa(\bm{0}) = \bm{0}$ and $(\cdot)_+ = \max \{0,\cdot\}$. As a shrinkage operator, it provides a simple closed-form solution for the $\bm{\theta}$-update (see \citet{boyd2010} for details).
\\

We derive the augmented Lagrangian function along with its first and second derivative w.r.t. $\bm{\beta}$ as
\begin{align*}
        \mathcal{L}(\bm{\beta}, \bm{\theta}, \bm{\nu}) &= f(\bm{\beta}) + g(\bm{\theta}) + \sum_{m=1}^M \left[\nu_m (\theta_m - \bm{K}_m \bm{\beta}) + \frac{\rho}{2} \|\theta_m - \bm{K}_m \bm{\beta} \|_2^2 \right] , \\
        \frac{\partial}{\partial \beta} \mathcal{L}(\bm{\beta}, \bm{\theta}, \bm{\nu}) &= f'(\bm{\beta}) + \sum_{m=1}^M \left[ -\nu_m  \bm{K}_m + \rho (-\theta_m + \bm{K}_m \bm{\beta}) \bm{K}_m  \right] =  U(\bm{\beta}) + [\rho (\bm{\beta}^T \bm{K}^T - \bm{\theta}^T) - \bm{\nu}^T] \bm{K}\\
        & = \bm{X}^T(\bm{\delta} - \hat{\bm{\mu}})  + [\rho (\bm{\beta}^T \bm{K}^T - \bm{\theta}^T) - \bm{\nu}^T]\bm{K}, \\
        \frac{\partial}{\partial \beta \partial \beta^T} \mathcal{L}(\bm{\beta}, \bm{\theta}, \bm{\nu}) &= f''(\bm{\beta}) + \sum_{m = 1}^M \left[ \rho \bm{K}_m^T \bm{K}_m \right] =  J(\bm{\beta}) + \rho \bm{K}^T \bm{K}  \\
        &= - \bm{X}^T \bm{W} \bm{X} + \rho \bm{K}^T\bm{K} ,
\end{align*}
with step size $\rho >0$, scaled dual variable $\bm{\nu} = (\nu_1,\dots,\nu_M)^T \in~\mathbb{R}^M$, score vector $U(\bm{\beta})$ as in (\ref{eq:LL-score}) and Hessian matrix $J(\bm{\beta})$ as in (\ref{eq:LL-hessian}).
Thus, by plugging-in both derivatives to the Newton-Raphson $\bm{\beta}$-updating step, our ADMM algorithm for the stacked regression parameter $\bm{\beta}~\in~\mathbb{R}^{PQ}$ in a multi-state model consists of the following steps:
\begin{enumerate}
\item Initialize $\bm{\beta}^0, \bm{\theta}^0$, and  $\bm{\nu}^0$.
\item Update until stopping criterion met:
    \begin{align*}
    \bm{\beta}^{r+1} &= \bm{\beta}^r + (\bm{X}^T \bm{W}^r \bm{X} + \rho \bm{K}^T\bm{K})^{-1} [ \bm{X}^T(\bm{\delta} - \hat{\bm{\mu}}^r)  + [\rho (\bm{\beta}^T \bm{K}^T - \bm{\theta}^T) - \bm{\nu}^T] K] ,\\
    \theta_m^{r+1} &= S_{\frac{\lambda_m w_m}{\rho}} (\bm{K}_m \bm{\beta}^{r+1} + \nu_m^r/\rho), ~m = 1,\dots,M ,\\
    \bm{\nu}^{r+1} &= \bm{\nu}^r + \rho(\bm{\theta}^{r+1} - \bm{K} \bm{\beta}^{r+1}) ,
    \end{align*}
\end{enumerate}
where parameter dimensions are $\bm{\theta}, \bm{\nu}~\in~\mathbb{R}^M$. $\lambda_m$ denotes the regularization parameters for the global lasso, fusion and group penalties, respectively, and $w_m =\sqrt{P}$ the group weights incorporating the group sizes corresponding to the group penalty. 
For the stopping criterion, we follow \citet{boyd2010} adapted for FSGL by \citet{beer_incorporating_2019} (Appendix Section 2.2) as
\begin{align*}
    &\|\bm{\theta}^{r+1}-K \bm{\beta}^{r+1}\|_2 < \epsilon_1   ~\text{and} \\
&\|\rho \bm{K}^T (\bm{\theta}^{r+1}-\bm{\theta}^{r})\|_2 < \epsilon_2 ,
\end{align*}
with $\epsilon_1 = \sqrt{PQ} \epsilon_{\text{abs}} + \epsilon_{\text{rel}} \max \lbrace \|\bm{K} \bm{\beta}^{r+1}\|_2,\|\bm{\theta}^{r+1}\|_2 \rbrace$, $\epsilon_2 = \sqrt{M} \epsilon_{\text{abs}} + \epsilon_{\text{rel}} \|\bm{K}^T \bm{\nu}^{r+1}\|_2$ and tolerances $\epsilon_{\text{abs}}, \epsilon_{\text{rel}}$ as in Subsection~\ref{sec:ADMM}. 
Regarding the ADMM step size $\rho > 0$, we follow \citet{beer_incorporating_2019} by implementing an adaptive step size proposed by \citet{he2000} to accelerate the convergence of the ADMM algorithm, i.e.
\begin{align*}
    \rho^{r+1} = 
    \left\{
   \begin{array}{ll}
      \tau \rho^r, & \text{if}~\|\bm{\theta}^{r+1}-K \bm{\beta}^{r+1}\|_2 > \eta \|\rho \bm{K}^T (\bm{\theta}^{r+1}-\bm{\theta}^{r})\|_2, \\
            \frac{\rho^r}{\tau}, & \text{if}~\|\bm{\theta}^{r+1}-K \bm{\beta}^{r+1}\|_2 < \eta \|\rho \bm{K}^T (\bm{\theta}^{r+1}-\bm{\theta}^{r})\|_2, \\
         \rho^r, & \text{otherwise},
   \end{array}
\right.
\end{align*}
where we set $\tau = 2$, $\eta =10$ and initialize $\rho^0 = 1$.
Algorithm~1 provides a summary of the adapted ADMM algorithm to FSGL penalized multi-state models (\textit{FSGLmstate}). \\
\begin{algorithm}
\centering
\caption{ADMM for fused sparse-group lasso penalized multi-state models (\textit{FSGLmstate})} 
	\begin{algorithmic}[1]
        \State Set $\bm{K} \in \mathbb{R}^{M \times PQ}$, $\alpha, \gamma \in [0,1]$, $\rho = 1$, $\epsilon_{\text{NR}} = 0.01$, and $v_{\text{NR}} =10^{-6}$.
        \State \textbf{initialize} $\bm{\beta}^0 = \bm{0}_{PQ}, \bm{\theta}^0 = \bm{0}_M, \bm{\nu}^0 = \bm{0}_M$.
        \Repeat 
            \State Update $\bm{\beta}^{r+1} = \bm{\beta}^r + (\bm{X}^T \bm{W}^r \bm{X} + \rho \bm{K}^T\bm{K})^{-1} [ \bm{X}^T(\bm{\delta} - \bm{\hat{\mu}}^r) + [\rho (\bm{\beta}^T \bm{K}^T - \bm{\theta}^T) - \bm{\nu}^T] \bm{K}]$,
            \State Update $\bm{\theta}_m^{r+1} = S_{\frac{\lambda_m w_m}{\rho}} (\bm{K_m} \bm{\beta}^{r+1} + \nu_m^r/\rho), ~m = 1,\dots,M$,
            \State Update $\bm{\nu}^{r+1} = \bm{\nu}^r + \rho(\bm{\theta}^{r+1} - K \bm{\beta}^{r+1})$,
        \Until $ \|\bm{\theta}^{r+1}-\bm{K} \bm{\beta}^{r+1}\|_2 < \epsilon_1$ and $\|\rho \bm{K}^T (\bm{\theta}^{r+1}-\bm{\theta}^{r})\|_2 < \epsilon_2$ for sufficiently small $\epsilon_1$ and $\epsilon_2$.
        \State \textbf{obtain} $\hat{\bm{\beta}} = \hat{\bm{\theta}}$.
	\end{algorithmic} 
\end{algorithm}

To tackle the dependency of the penalized estimation solution on relative variable scales, standardization is performed for continuous covariates before applying penalization, i.e. $x^*_{p.q} = \frac{x_{p.q}}{\hat{\sigma}_{x_{p.q}}}$, where $\hat{\sigma}_{x_{p.q}}$ denotes the empirical standard deviation of $x_{p.q}$. For interpretion, the regression coefficients have to be scaled back after estimation. \\

The algorithm can be easily amended to situations in which certain covariates should not be penalized (e.g. established clinical predictors). Therefore, we introduce an individual penalty scaling factor $\zeta_m \geq 0, m = 1,\dots,PQ$, which allows different penalties for each variable, i.e. $\lambda_m = \lambda \zeta_m$.\citep{friedman2010regularization}
Unpenalized parameters get a penalty scaling factor set to zero, i.e. $\zeta_m = 0$ for $m \in \lbrace 1,\dots,PQ \rbrace$. \\

Further, it is important to note that the ADMM algorithm does not generate exact zeros for the $\hat{\bm{\beta}}$-solution.\cite{andrade2021, parka2022admm} However, the estimated auxiliary variable $\hat{\bm{\theta}}$ is sparse, so that variable selection results are based on the derived estimate $\hat{\bm{\theta}}$. Thus, we get the final estimated penalized regression parameters as $\hat{\bm{\beta}} = \hat{\bm{\theta}}$.

\subsection{Selection of tuning parameters}\label{sec:Tuning}

For tuning parameter selection, we focus on the approximate \textit{generalized cross-validation} (GCV) statistic\citep{cravenwahba1978}. This selection criterion was used by \citet{tibshirani2005fused} for the fused lasso and \citet{fanli2002} for variable selection in penalized Cox models. GCV is an estimator of the predictive ability of a model\citep{jansen2015gcv}, which is defined as
    \begin{align*}
        \text{GCV}(\lambda) = \frac{L(\hat{\bm{\beta}})}{N[1-e(\lambda)/N]^2},
    \end{align*}
where $\lambda$ is a general tuning parameter. The effective number of model parameters for the Cox proportional hazards model in the last step of the Newton-Raphson algorithm iteration\citep{fanli2002} is approximated as 
\begin{align*}
    e(\lambda) = \text{tr}\left[\left\{\dfrac{\partial^2}{\partial \beta \partial \beta^T} L(\hat{\bm{\beta}}) + \Sigma_\lambda(\hat{\bm{\beta}})\right\}^{-1} \dfrac{\partial^2}{\partial \beta \partial \beta^T} L(\hat{\bm{\beta}})\right],
\end{align*} with 
\begin{align*}
\Sigma_\lambda(\hat{\bm{\beta}}) &= \text{diag}\left\{\dfrac{p'(\hat{\beta}_{1,1})}{|\hat{\beta}_{1,1}|},\dots, \dfrac{p'(\hat{\beta}_{P,Q})}{|\hat{\beta}_{P,Q}|} \right\}
\end{align*}
and $p'(\cdot)$ denoting the first derivative of the locally quadratic approximated penalty function.
The optimal tuning parameter is then selected as $\lambda_{\text{opt}}~=~\text{arg min}_\lambda \{ \text{GCV}(\lambda)\}$. For the selection of an optimal combination of multiple tuning parameters, grid search\citep{tibshirani2005fused} along with the Brent optimization algorithm\citep{brent1973} is utilized.

%%%%%%%%%%%%%%%%%%%%%%%%%%%%%%%%%%%%%%%%%%
%\newpage
\section{Simulation study}\label{sec:simulations}
This Section describes the design of a proof-of-concept simulation study for evaluating FSGL penalized multi-state models in terms of variable selection in Subsection~\ref{sec:sim_design} and illustrates corresponding results in Subsection~\ref{sec:sim_results}.

\subsection{Simulation design}\label{sec:sim_design}

The aim of the following proof-of-concept simulation study is to evaluate the variable selection procedure based on FSGL penalized multi-state models in terms of its ability to select a sparse model distinguishing between relevant transition-specific effects and equal cross-transition effects. As a methodological phase II simulation study, it offers empirical evidence to demonstrate validity in finite samples across a limited range of scenarios.\citep{heinze2024phases} The corresponding ADEMP criteria of the simulation study are summarized in Table~\ref{tab:ADEMP}. A detailed simulation study plan according to ADEMP-PreReg\citep{siepe_ademp} can be found in the Supporting Information.

\begin{table}[b]
\centering
\caption{ADEMP criteria of the simulation study according to \citet{morris_using_2019}.}
\label{tab:ADEMP}
\begin{tabularx}{\textwidth}{ll}\hline
\toprule
\textbf{ADEMP criterion} & \textbf{Definition}  \\ \hline\hline
Aim & Evaluation of sparse variable selection detecting relevant transition-specific effects and \\
 & equal cross-transitions effects\\
Data-generating mechanism & Multi-state model based on transition-specific hazards models \\
Estimand/target & Regression coefficients \\
Methods & Unpenalized Cox-type multi-state estimation with ADMM optimization; \\
& Lasso penalized multi-state model with ADMM optimization (LASSOmstate); \\
& Fused sparse-group lasso penalized multi-state model with ADMM optimization  (FSGLmstate)\\
Performance measures & True positive rate (TPR); False discovery rate (FDR);\\
& Bias; Mean squared error (MSE) \\
\bottomrule \hline
\end{tabularx}
\end{table}

\vspace{-.4cm}
\subsubsection*{Data-generating mechanism}\label{sec:DGM}

In each simulation run, we generate multi-state data with a sample size of $N=1000$ from the AML multi-state model displayed in Figure~\ref{fig:diagram_AML} based on transition-specific hazards regression as a nested series of competing risks experiments according to \citet{beyersmann_simulating_2009} 
Thus, data has been generated by the following data-generating process:
Waiting times in state $l$ are generated from an exponential distribution with hazards $h_{l\cdot} = \sum_{k=1, k \neq l}^9 h_{lk}$, $l=1,\dots,9$. Transition-specific baseline hazards are set constant to $h_{0,q}(t)~=~0.05$ for all transitions $q~=~1,\dots,8$. We synthesize two independent biomarkers as binary covariates $X_{p,i} \sim \mathcal{B}(0.5), ~p~=~1,2, i~=~1,\dots,1000$. The true regression parameters for biomarker $X_1$ are set to $\bm{\beta}_{1,1} = 1.5$ for transition~1, $\beta_{1,3} = \beta_{1,7} = 1.2$ for transitions~3 and 7, $\beta_{1,4} = \beta_{1,8} = -0.8$ for transitions~4 and 8 and $\beta_{1,2} = \beta_{1,5} = \beta_{1,6} = 0$ for transitions~2, 5 and 6. Similar transitions are 3 and 7, i.e. from first complete remission (CR1) to first relapse and from second complete remission (CR2) to second relapse, as well as 4 and 8, i.e. CR1 to death in CR1 and CR2 to death in CR2. Thus, covariate $X_1$ has equal effects on these two pairs of similar transitions. Covariate $X_2$ has no effect on any transition, i.e. $\beta_{2,1} = \dots = \beta_{2,8} = 0$.

\vspace{-.4cm}
\subsubsection*{Target}\label{sec:target}

Our primary target focuses on the true non-zero regression coefficients $\beta_{p.q}$ from the penalized multi-state Cox-type proportional hazards models
\begin{align*}
    h_q (t|x) = h_{0,q}(t) \exp \lbrace  \bm{\beta}_q^T \bm{x} \rbrace , ~ q=1,\dots,8,
    \label{hazard-model}
\end{align*}
where $h_{0,q}(t)$ denotes the baseline hazard rate of transition $q$ at time $t$, $\bm{x}=(x_1,\dots,x_P)^T \in \mathbb{R}^P$ the vector of covariates and $\bm{\beta}_q \in \mathbb{R}^P$ the vector of transition-specific regression coefficients for $P$ covariates.

\vspace{-.4cm}
\subsubsection*{Methods}\label{sec:method}
We aim to compare the \textit{FSGLmstate} algorithm to unpenalized multi-state Cox-type estimation and lasso penalized estimation (\textit{LASSOmstate}) based on ADMM optimization.
For fitting Cox-type multi-state models by ADMM optimization as described in Section~\ref{sec:ADMM-mstate}, we chose the following parameter settings: The ADMM variables are initialized as $\bm{\beta}^0 = \bm{\theta}^0 = \bm{\nu}^0 = \bm{0}$ and the adaptive ADMM step size as $\rho^0 = 1$. The step size in gradient descent is set to $\epsilon_{\text{GD}} = 0.01$, the tolerance of the stopping criterion for Cox estimation $\text{tol}_{\text{GD}}~=~10^{-6}$, the relative and absolute tolerance for the ADMM stopping criterion $\epsilon_{\text{rel}}~=~10^{-2}$ and $\epsilon_{\text{abs}}~=~10^{-4}$ and the maximum number of iterations to $\text{max}_{\text{iter}} = 500$.
For each combination of tuning parameters $\alpha, \gamma \in \{0, 0.25, 0.5, 0.75, 1\}$, the optimal overall penalty parameter $\hat{\lambda}_{\text{opt}} >0$ is selected by minimal GCV over a grid of $\lambda \in \{0.01,\dots,500\}$, equally spaced on a logarithmic scale.

\vspace{-.4cm}
\subsubsection*{Performance measures}\label{sec:performance_measures}

Regularization performance is assessed by true positive rates (TPR) and false discovery rates (FDR) of variable selection. Median counts of true positives (TP), true negatives (TN), false positives (FP) and false negatives (FN) of variables over all simulations are calculated. Based on these absolute counts, TPR is calculated as $\text{TPR} = \frac{\text{TP}}{\text{TP+FN}}$. Further, FDR is defined as the number of unrelated variables selected (i.e. false positives) divided by the total number of selected variables, i.e. $\text{FDR} = \frac{\text{FP}}{\text{TP+FP}}$. 

For quantifying the estimation bias, $\text{Bias}(\hat{\bm{\beta}}) = \hat{\bm{\beta}} - \bm{\beta}$, for the non-zero covariates, the mean squared error (MSE) over all simulation iterations is used. The MSE for the non-zero covariates is defined as
\begin{align*}
    \text{MSE}_{nz}(\hat{\bm{\beta}}) = \frac{1}{d} \sum_{p,q:\beta_{p,q} \neq 0} (\hat{\beta}_{p,q} - \beta_{p,q} )^2 , 
\end{align*}
where $d$ denotes the number of non-zero covariates with $\beta_{p,q} \neq 0$ of the true model. The mean bias and mean MSE averaged over the non-zero predictors over all simulation runs along with Monte Carlo standard errors (MCSE) are calculated according to \citet{morris_using_2019}.

The number of simulation runs is based on the TPR as one of the primary performance measures of interest. Thus, we need $n_{\text{sim}}~=~225$ simulation repetitions per scenario as we aim for TPR $\geq~0.9$ and MCSE(TPR) $\leq~0.02$ and assume $\text{MCSE}(\widehat{\text{TPR}}) \leq~0.15$, resulting in $n_{\text{sim}}~=~\frac{0.9 \cdot 0.1}{0.02^2} = 225$ and $n_{\text{sim}}~=~\frac{0.15^2}{0.02^2} = 225$, respectively.

\subsection{Simulation results}\label{sec:sim_results}
\begin{figure}[tb]
\centering
\includegraphics[width=14cm]{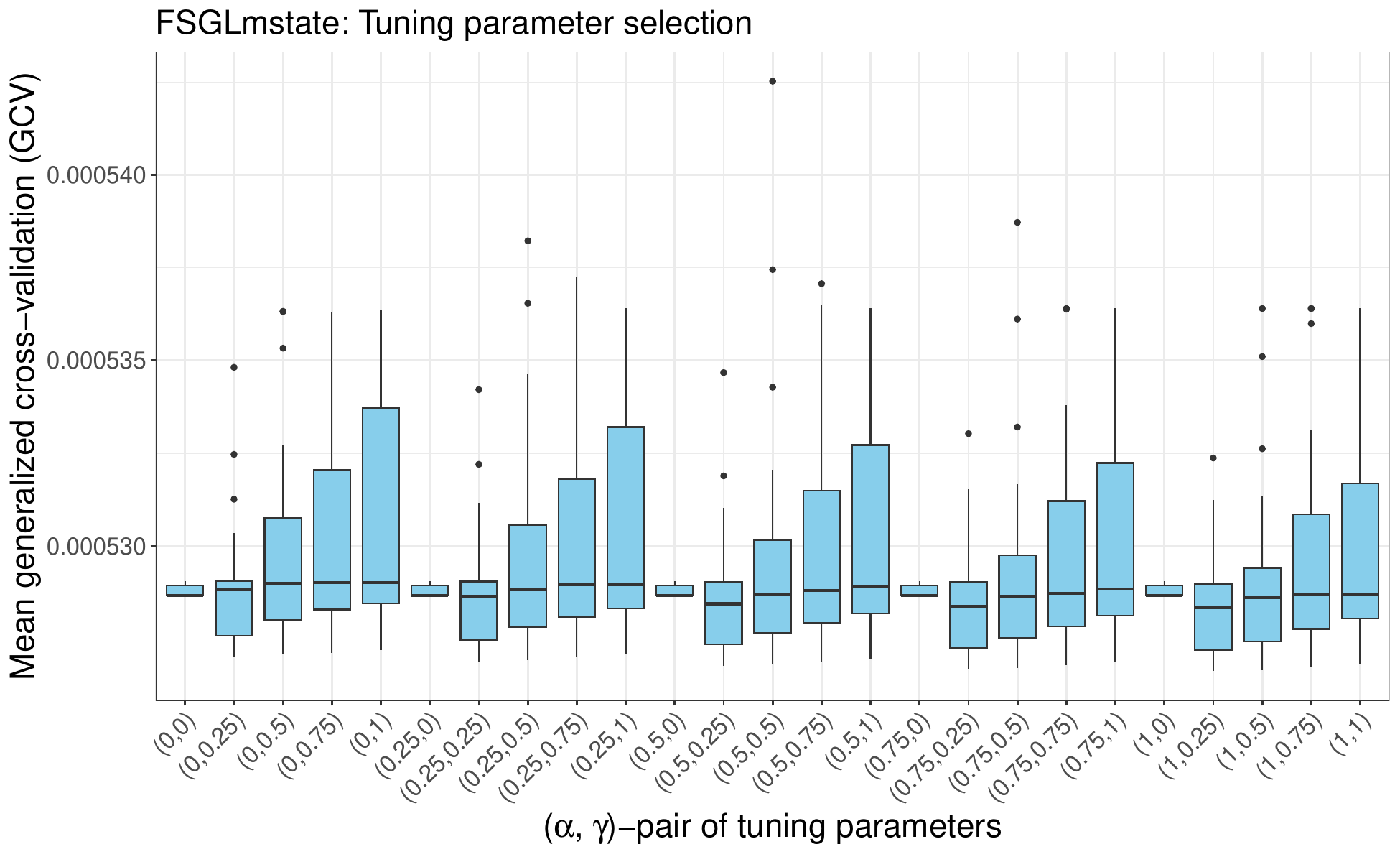}
\caption{Tuning parameter selection results for FSGLmstate: Mean generalized cross-validation (GCV) statistics across all pre-selected combinations of penalty parameters $(\alpha,\gamma)$ over all simulation runs. The pair $(\alpha, \gamma) = (1, 1)$ corresponds to the global lasso penalty.} %Dots illustrate mean GCV for each $\lambda$ over the simulated data sets.}
\label{fig:meanGCV_nobs1000}
\end{figure}

\begin{figure}[!h]
\centering
\includegraphics[width=17cm]{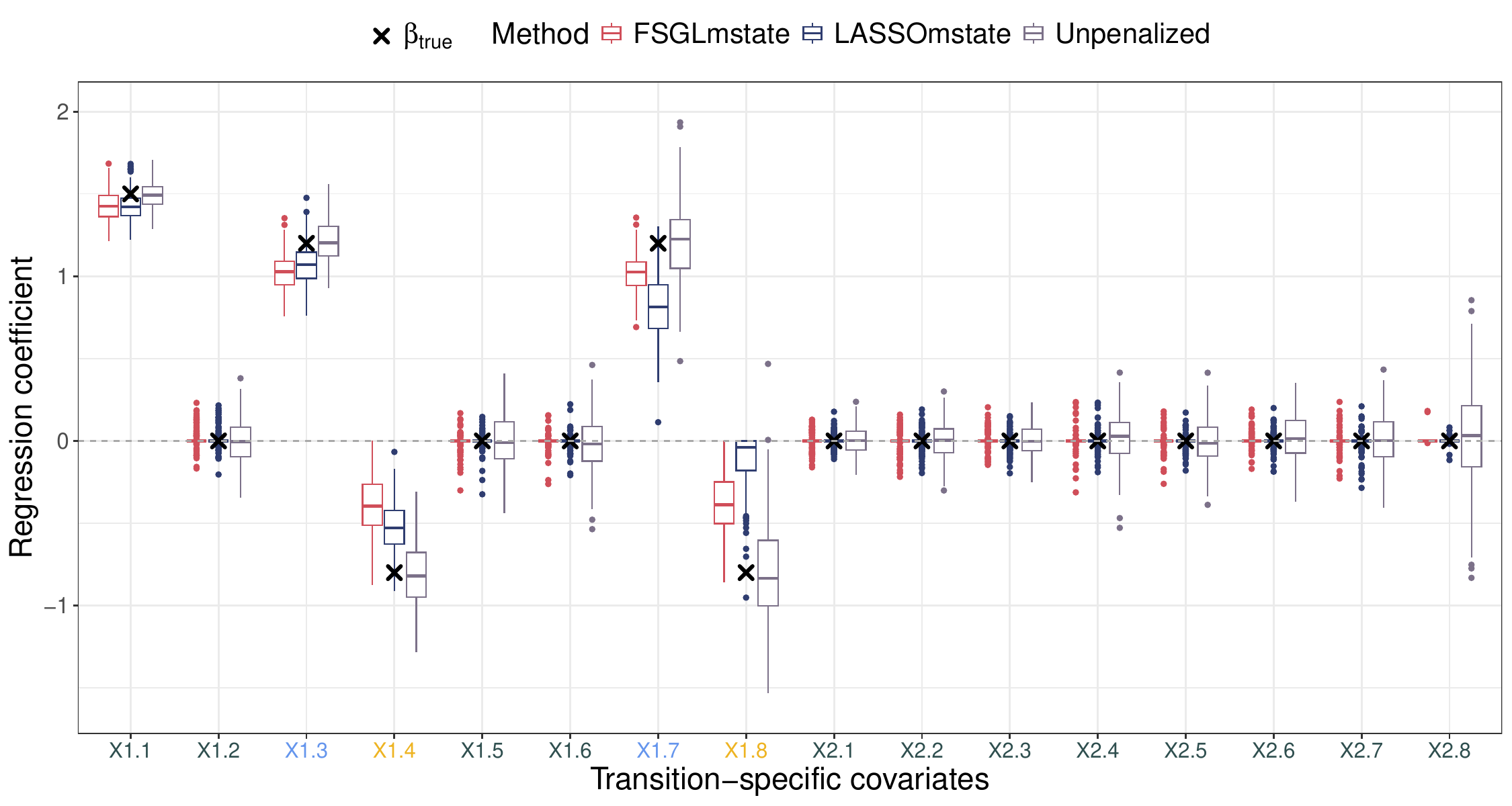}
\caption{Boxplots of estimated regression coefficients based on simulated data of the 9-state AML model with eight transitions and two binary covariates. $X1.3$ and $X1.7$ as well as $X1.4$ and $X1.8$ refer to transitions with true equal effects of covariate $X_1$. Covariate $X_2$ has no true effect on any transition. Dots depict estimated covariate effects based on $\hat{\lambda}_{\text{opt,L}}$ and $\hat{\lambda}_{\text{opt,FSGL}}$ of each simulated data set. True underlying covariate effects $\bm{\beta}_{\text{true}}$ are denoted as crosses ($\times$).}
\label{fig:boxplots_coef_nobs1000}
\end{figure}

\begin{figure}[tb]
\centering
\begin{subfigure}[b]{0.49\textwidth}
    \centering
    \includegraphics[width=\textwidth]{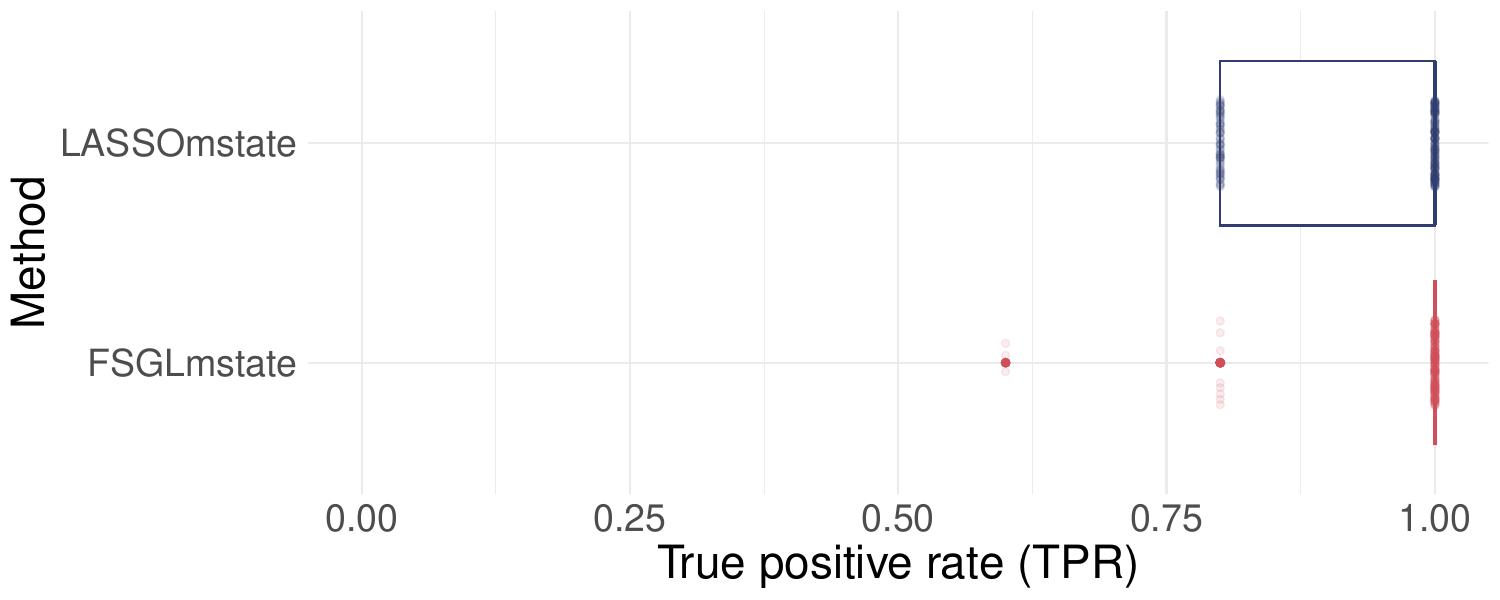}
    %\caption{True Positive Rates (TPR)}
    %\label{fig:TPR_coef_nobs1000}
\end{subfigure}
\hfill
\begin{subfigure}[b]{0.49\textwidth}
    \centering
    \includegraphics[width=\textwidth]{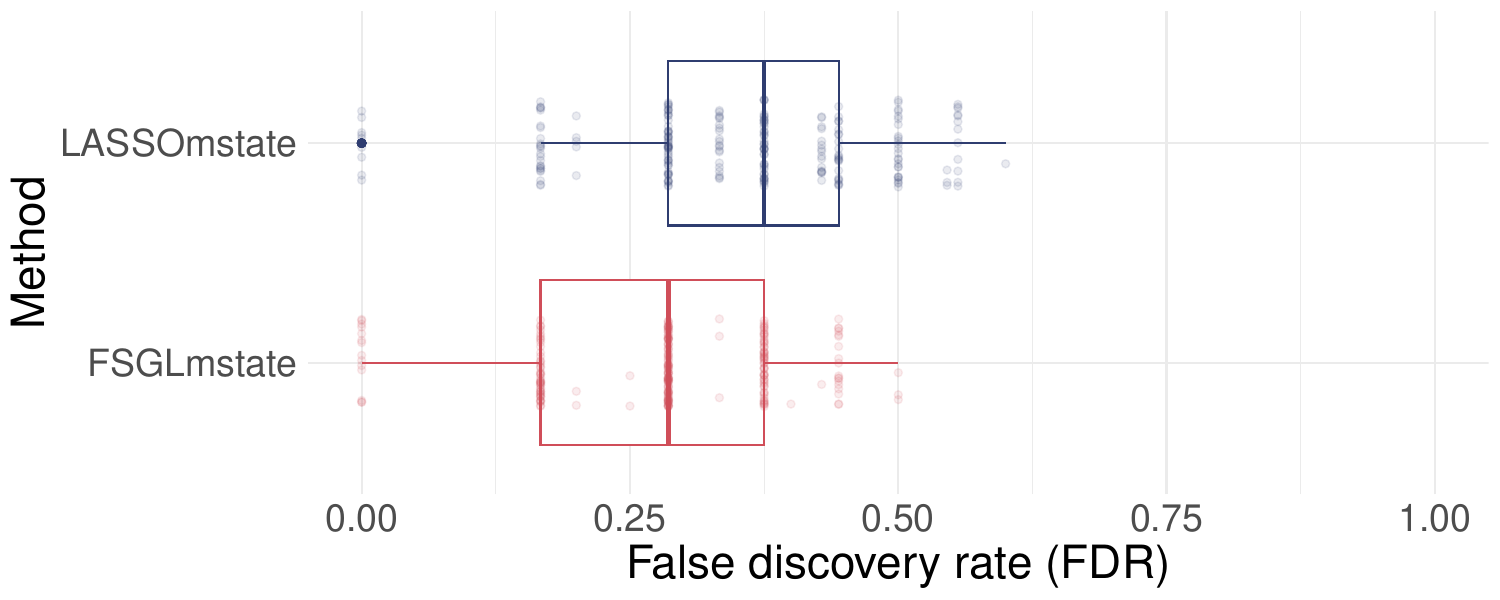}
    %\caption{False Discovery Rates (FDR)}
    %\label{fig:FDR_coef_nobs1000}
\end{subfigure}
\caption{Variable selection results in terms of true positive rates (TPR) and false discovery rates (FDR) for LASSOmstate and FSGLmstate. Dots illustrate TPR and FDR of each simulated data set.}
\label{fig:combined_plot_TPR-FDR}
\end{figure}

\begin{figure}[t]
\centering
\begin{subfigure}[b]{0.65\textwidth}
    \centering
    \includegraphics[width=\textwidth]{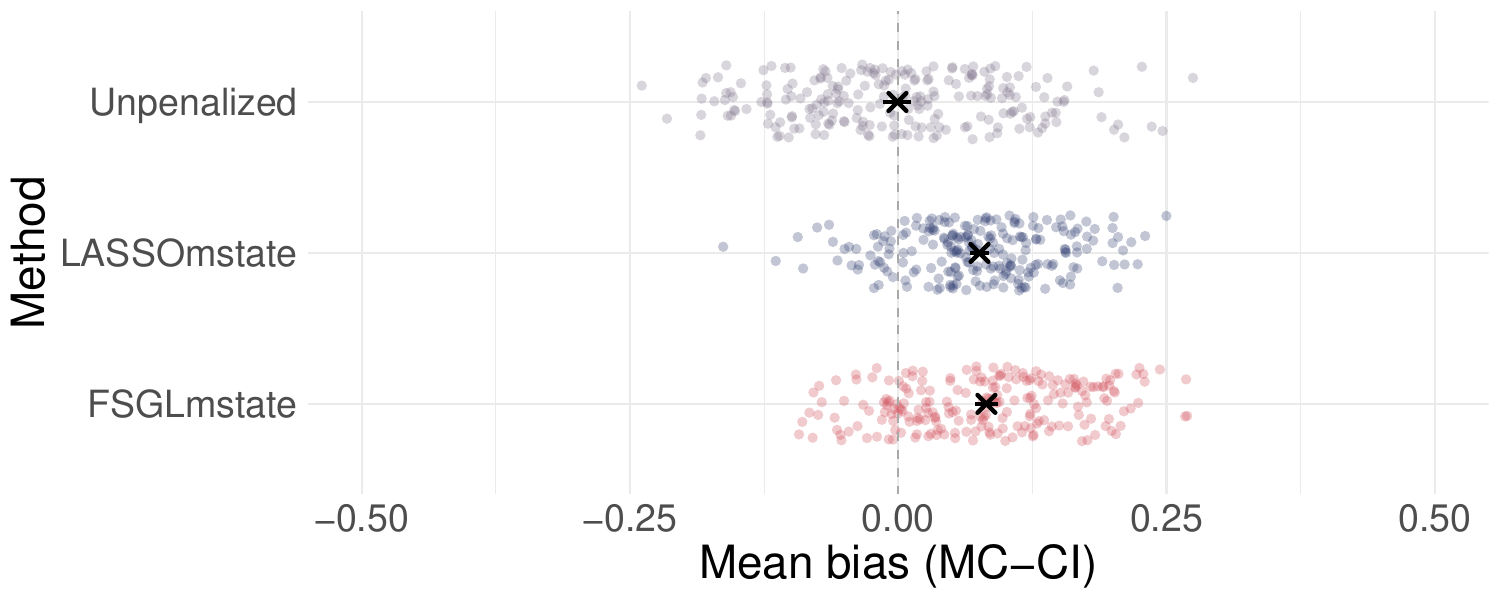}
\end{subfigure}
\hfill
\begin{subfigure}[b]{0.65\textwidth}
    \centering
    \includegraphics[width=\textwidth]{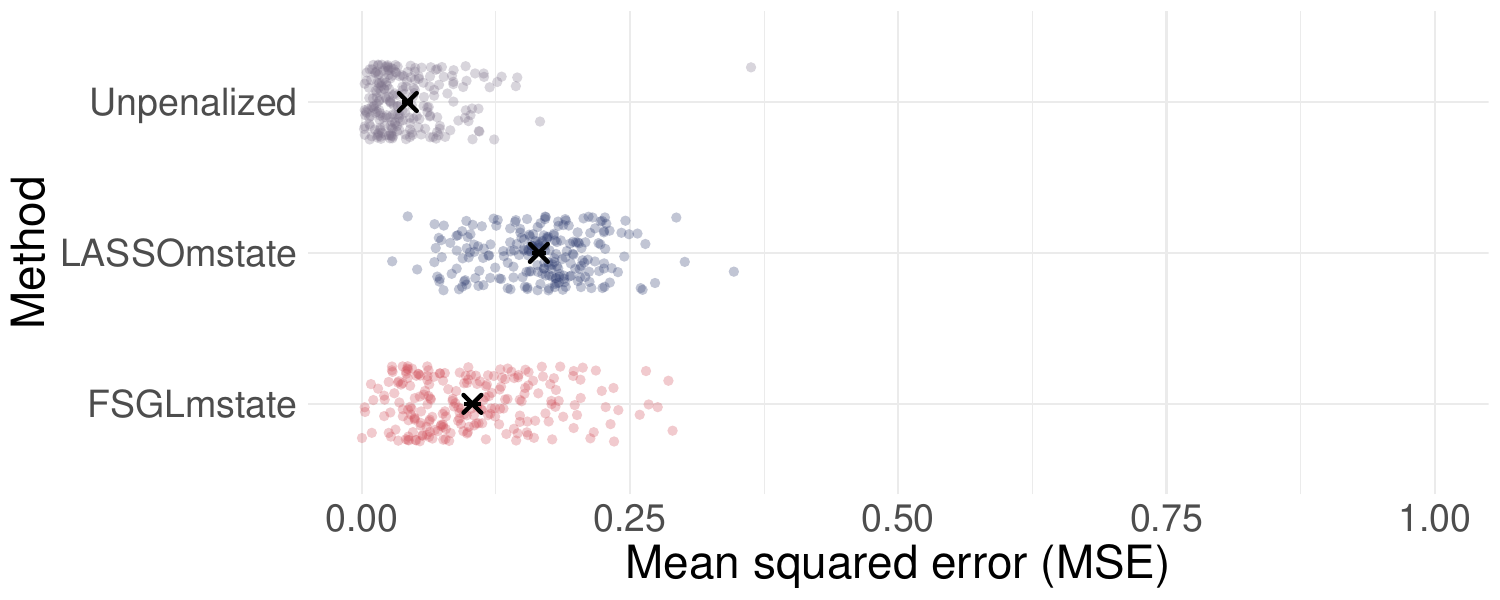}
\end{subfigure}
\caption{Mean bias and mean squared error (MSE) of estimating the non-zero covariate effects along with 95\% Monte Carlo confidence intervals (MC-CI). Dots illustrate mean bias and mean MSE of a single simulated data set.}
\label{fig:combined_plot_bias-mse}
\end{figure}

This Section summarizes the main simulation findings, with full results available in the Supporting Information. 
Tuning parameter selection by minimal GCV is illustrated in Figure~\ref{fig:meanGCV_nobs1000} for FSGLmstate. Boxplots depict mean GCV across combinations of tuning parameter pairs $(\alpha,\gamma)$ for a grid of penalty parameter $\lambda \in~\{0.01,\dots,500\}$ over all $n_{\text{sim}}=225$ simulated data sets. For LASSOmstate corresponding to $(\alpha, \gamma) = (1,1)$, the most frequent lowest GCV is obtained for the optimal tuning parameter $\hat{\lambda}_{\text{opt,L}}=8.6$ with mean $\text{GCV}(\hat{\lambda}_{\text{opt,L}})\cdot 1000 = 0.52597$ over all simulations. For FSGLmstate, the tuning parameter combination $(\alpha, \gamma) = (1,0.25)$ yields the most frequent lowest GCV for $\hat{\lambda}_{\text{opt,FSGL}}=38.1$ with mean $\text{GCV}(\hat{\lambda}_{\text{opt,FSGL}})\cdot 1000 = 0.52663$ over all simulated data sets with the corresponding penalty parameter combination.  
The regularization performance of the FSGLmstate algorithm in comparison to unpenalized and lasso penalized multi-state Cox-type estimation is depicted in Figure~\ref{fig:boxplots_coef_nobs1000}. For our simulation setting with $N=1000$ observations and $PQ=16$ regression parameters, unpenalized Cox-type estimation serves as a gold standard.
The boxplots illustrate the estimated regression coefficients of the binary covariates based on $\hat{\lambda}_{\text{opt,L}}$ and $\hat{\lambda}_{\text{opt,FSGL}}$. Whereas LASSOmstate identifies the non-zero effects of $\beta_{1,1} = 1.5$, $\beta_{1,3} = \beta_{1,7} = 1.2$ and $\beta_{1,4}~=~-0.8$, the negative effect of $\beta_{1.8} = -0.8$ for the late transition 8 from CR2 to death in CR2 is set to zero on average. FSGLmstate recognizes the similarity structure of the covariate effect pairs $\beta_{1,3} = \beta_{1,7} = 1.2$ as well as $\beta_{1,4} = \beta_{1.8} = -0.8$ while setting all other true negative covariate effects to zero. The unpenalized Cox-type estimation based on ADMM optimization identifies all non-zero effects, but inherently does not perform regularization, which results in larger variance for all true negative coefficients.
Figure~\ref{fig:combined_plot_TPR-FDR} depicts variable selection results in terms of TPR and FDR for LASSOmstate and FSGLmstate. Whereas FSGLmstate more often detects all non-zero regression effects, LASSOmstate's estimated TPR varies between 0.8 and 1.0 (left panel). With regard to FDR, FSGLmstate has a median estimated FDR of 0.29 and LASSOmstate of 0.38 (right panel). 
Figure~\ref{fig:combined_plot_bias-mse} illustrate the mean bias and MSE of estimating the non-zero covariate effects along with MCSE. As expected, unpenalized Cox-type estimation exhibits smallest mean bias and MSE of estimating the non-zero covariates in our simulation setting with $N=1000$ observations. Notably, FSGLmstate provides smaller mean MSEs than LASSOmstate.

%%%%%%%%%%%%%%%%%%%%%%%%%%%%%%%%%%%%%%%%%%
%\clearpage
\section{Application to leukemia data}\label{sec:application}

The potential of FSGL penalized multi-state models is further investigated in an illustrative application to leukemia data.
The AMLSG 09-09 study is a randomized phase III trial conducted between 2010 and 2017 at 56 study hospitals in Germany and Austria. The clinical trial evaluated intensive chemotherapy with or without gemtuzumab ozogamicin (GO) in patients with NPM1-mutated AML. Final analysis results for the single and composite endpoints event-free survival (EFS), overall survival (OS), CR rates and cumulative incidence of relapse (CIR) with long-term follow-up are published in \citet{dohner2023-09-09}. In conclusion, primary endpoints of the trial in terms of EFS and OS were not met. Additional gene mutation data are available for $N=568$ study patients.

\begin{figure}[b]
\centering
\includegraphics[width=17cm]{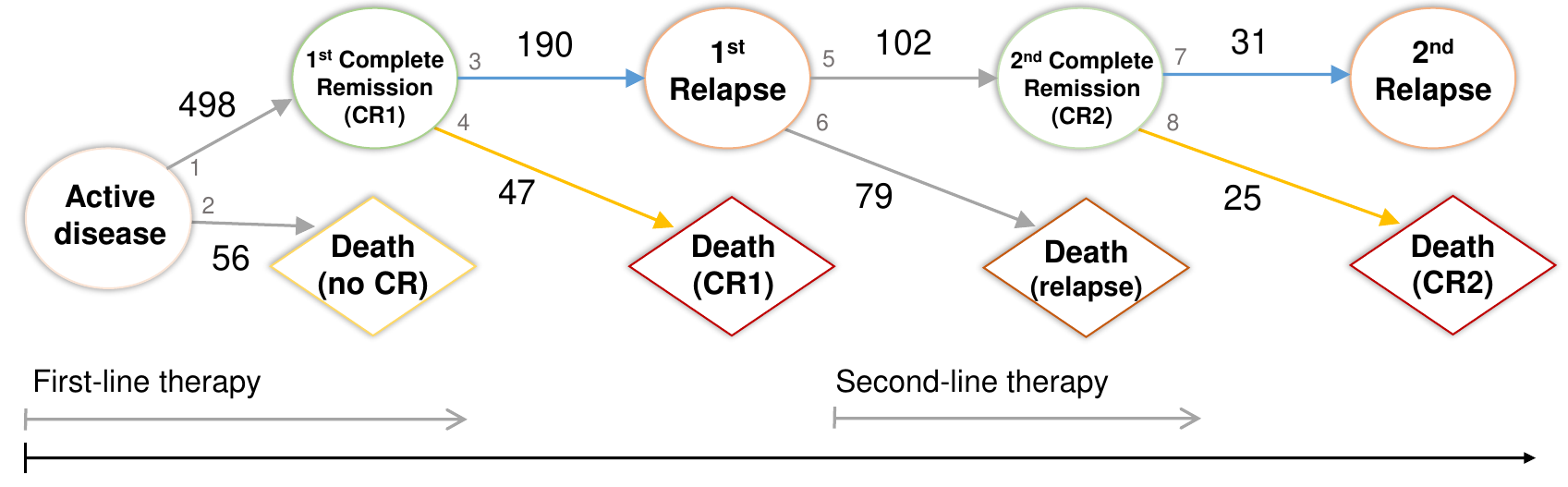}
\caption{Event counts of the multi-state model for acute myeloid leukemia (AML) with nine states and eight transitions based on the AMLSG 09-09 trial data. Gray numbers indicate the corresponding transition.}
\label{fig:diagram_events_AML-09-09}
\end{figure}

\begin{figure}[b]
\centering
\includegraphics[width=13cm]{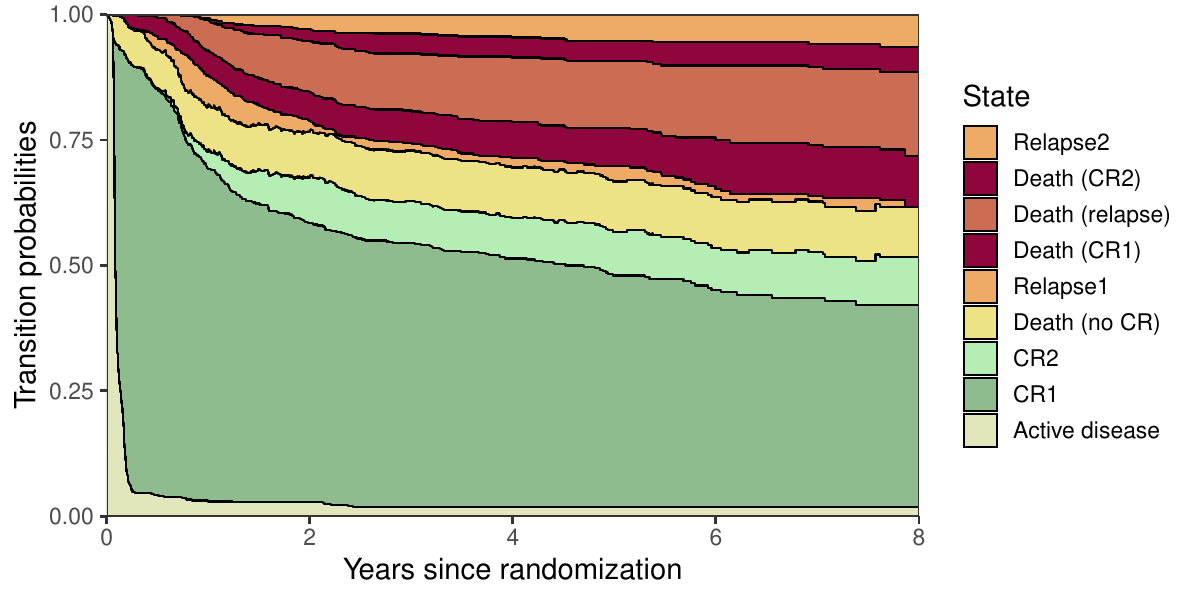}
\caption{Stacked transition probabilities to all states from randomization derived from the multi-state model for acute myeloid leukemia (AML) based on the AMLSG 09-09 trial data. The distance between two adjacent curves represents the probability of being in the corresponding state. CR: Complete remission.}
\label{fig:09-09_cum-haz}
\end{figure}

Our motivating 9-state model for AML along with event counts based on the 09-09 trial data is illustrated in Figure~\ref{fig:diagram_events_AML-09-09}. Late transitions 7 and 8 are rather rarely observed with few events ($E_7 = 31, E_8 = 25$). Derived from this multi-state model, Figure~\ref{fig:09-09_cum-haz} depicts the stacked transition probabilities to all states from randomization. The probability of being in an intermediate state can fluctuate over time, either increasing or decreasing, while the absorbing state probabilities can only increase over time.%\citep{bakunina_2022}

\begin{figure}[b]
\centering
\includegraphics[width=18cm]{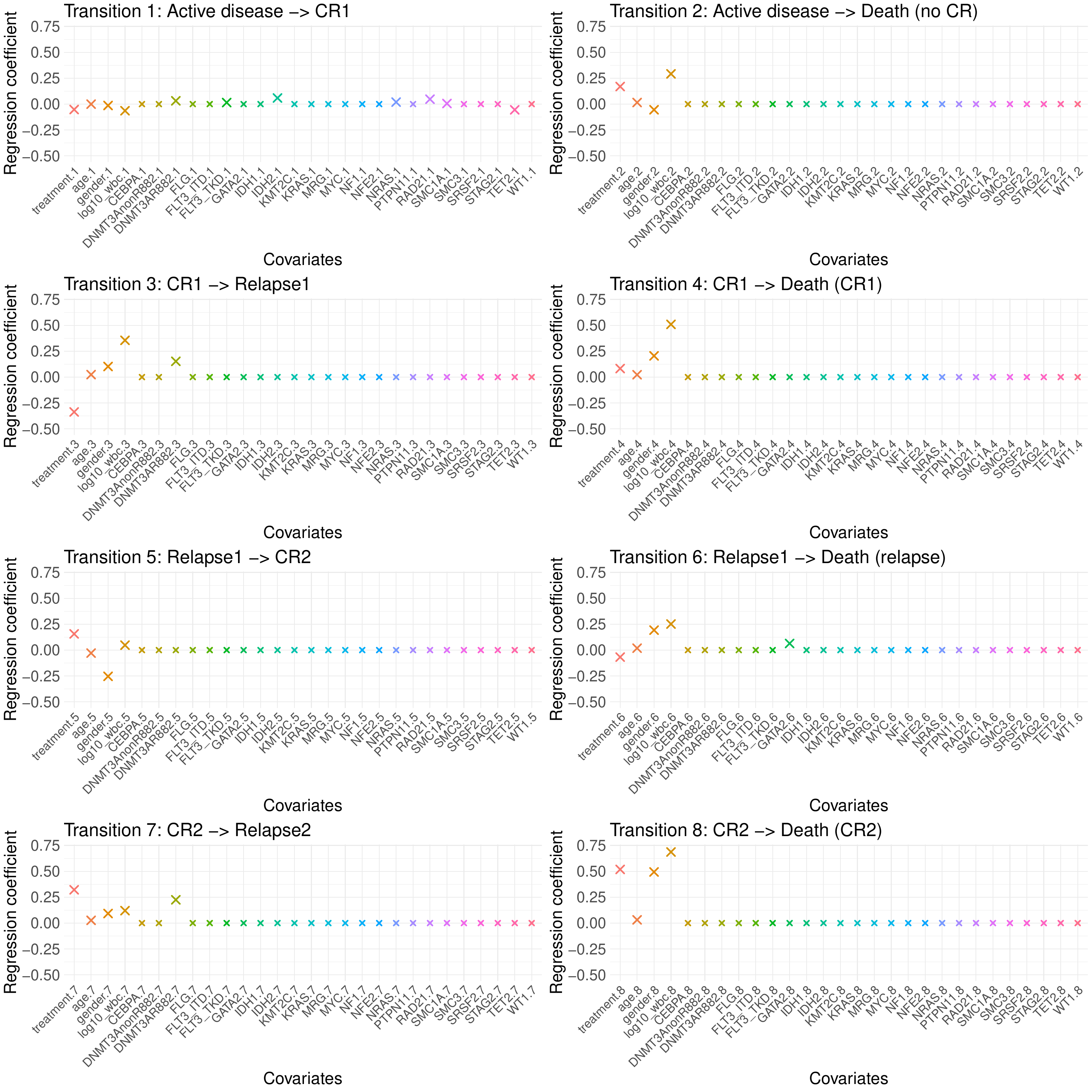}
\caption{Estimated regression effects of clinical and mutation variables by FSGLmstate separately for each transition derived from the 9-state model for acute myeloid leukemia (AML) based on the AMLSG 09-09 trial data. Larger crosses ($\times$) depict non-zero effects.}
\label{fig:09-09_coef_FSGL_all-trans}
\end{figure}

For our 9-state model, we investigate covariate effects of $P=24$ gene mutations with a prevalence of >3\% along with the $P_c~=~4$ established clinical predictors treatment  (GO vs. standard), age (years), sex (male vs. female) and $\log_{10}$ transformed white blood cell count ($10^9$ cells/l). Considering these $P=28$ covariates and $Q=8$ transitions, we need to incorporate $(P+P_c)Q~=~28~\cdot~8~=~224$ regression parameters. The clinical predictors should persist unpenalized, thus we apply the FSGL penalty to the remaining 192 mutation parameters. We assume similarity for transitions 3 and 7, i.e. CR1 to first relapse and CR2 to second relapse, as well as transitions 4 and 8, i.e. CR1 to death in CR1 and CR2 to death in CR2, so that we have $s=2$ pairs of similar transitions.
With respect to a-priori expert knowledge on similarity and grouping structures in AML mutations, tuning parameter combinations are investigated for $\alpha \in \{0.5, 0.75, 1 \}$ with more weight on the global lasso and $\gamma \in \{0, 0.25, 0.5\}$ putting more weight to the fusion penalty. Among all pre-defined pairs $(\alpha, \gamma)$, the optimal combination of regularization parameters  $(\hat{\alpha}_{\text{opt,FSGL}}, \hat{\gamma}_{\text{opt,FSGL}}) = (0.75, 0.5)$ and $\hat{\lambda}_{\text{opt,FSGL}} = 20$ is then selected by minimal GCV over the grid $\lambda \in \{ 0.01, \dots, 500\}$.
Figure~\ref{fig:09-09_coef_FSGL_all-trans} depicts all estimated regression coefficients of clinical and mutation variables by FSGLmstate separately for each transition.
In consistence with final analysis results for CIR, treatment has a negative regression effect on transition 3, i.e. from CR1 to first relapse, suggesting an anti-leukaemic efficacy of intensive chemotherapy including GO ($\hat{\beta}_{\text{treatment.3}} = -0.34$). With respect to molecular markers, mutations of the DNA methylation gene \textit{DNMT3A}\textsuperscript{R882} are selected for transition 3 from CR1 to first relapse, as well as for transition 7 from CR2 to second relapse. This result aligns with accompanying gene mutation analyses of \citet{cocciardi2024}, where \textit{DNMT3A}\textsuperscript{R882} mutations were associated with an increased CIR.
%cite Cocciardi et al. (2024) when available

%%%%%%%%%%%%%%%%%%%%%%%%%%%%%%%%%%%%%%%%%%
\section{Discussion}\label{sec:discussion}
In this paper, we propose FSGL penalized multi-state models for data-driven variable selection and dimension reduction in order to capture pathogenic disease processes more accurately while incorporating clinical and molecular data. The objective was to select a sparse model based on high-dimensional molecular data by extended regularization methods. We adapted the ADMM algorithm to FSGL penalized multi-state models combining the penalization concepts of general sparsity, pairwise differences of covariate effects along with transition grouping. 
Thus, FSGL penalized multi-state models tackle sparse model building while incorporating a-priori information about the covariate and transition structure into a prediction model. Further, the ADMM algorithm can quite efficiently handle large-scale problems due to the decomposability of the objective function as well as superior convergence properties.

The proof-of-concept simulation study evaluated the FSGLmstate algorithm's regularization performance to select a sparse model incorporating only relevant transition-specific effects and similar cross-transition effects. Compared to unpenalized and global lasso penalized estimation, FSGLmstate identifies similarity and grouping structures depending on the choices of the corresponding tuning parameters.

The real-world data application on a phase III AML trial illustrated the utility of an FSGL penalized multi-state model to reduce model complexity while combining clinical and molecular data. Whereas an unpenalized 9-state model incorporating all established clinical predictors along with high-dimensional mutation information based on the study data suffers from overfitting due to few events per variable, our FSGLmstate approach allows to fit a penalized 9-state model combining clinical predictors and mutation variables.

Several improvements and extensions of the proposed FSGL penalty to multi-state models offer further research directions. One limitation of our work is that time-dependent covariates, e.g. allogeneic stem cell transplantation, and time-dependent effects are not yet incorporated. Further, post-selection inference requires to be investigated. Besides, the algorithm needs further adaptations to enhance computational speed and efficiently handle very high dimensions with $P \gg N$. Additionally, different tuning parameter selection criteria should be investigated and extensive phase III simulations for empirical method comparisons are required to evaluate the performance of our variable selection method across a wide range of settings.

%%%%%%%%%%%%%%%%%%%%%%%%%%%%%%%%%%%%%%%%%%
\section*{Computational details}

All implementations and statistical analyses are performed utilizing the statistical computing language R, version 4.4.1 \citep{R}, along with the R~packages \texttt{mstate} \citep{wreede_mstate_2011}, \texttt{penalized}\citep{goeman2010} and \texttt{penMSM}\citep{penMSM} among others.
R~code to reproduce simulation study results and manuscript figures is available in the Supporting Information.

\section*{Acknowledgements}
%\begin{acks}
    The authors would like to thank Maral Saadati, Jan Beyersmann and Jörg Rahnenführer for fruitful discussions. Further, we are grateful to the German-Austrian AML study group (AMLSG) for providing the 09-09 trial data, as well as all patients and centers involved in the study.
    The work of the corresponding author was supported by the Deutsche Forschungsgemeinschaft (DFG, German Research Foundation, grant number 514653984).
%\end{acks}

\section*{Conflict of Interest}
    The authors have declared no conflict of interest.

%\subsection*{References}
%\bibliography{wileyNJD-APA}
\bibliography{mybib}

\end{document}